# Evaluation of Mother-Daughter Architectures for Asteroid Belt Exploration

Leonard D. Vance,[1]
*Space and Terrestrial Robotic Exploration Laboratory, University of Arizona, Tucson Arizona, 85721 USA*

Jekan Thangavelautham[2]
*Space and Terrestrial Robotic Exploration Laboratory, University of Arizona, Tucson Arizona, 85721 USA*

and

Erik Asphaug[3]
*Lunar and Planetary Laboratory, University of Arizona, Tucson Arizona, 85721 USA*

**This paper examines the effectiveness of an asteroid exploration architecture comprised of multiple nanosatellite sized spacecraft deployed from a single mother ship into a heliocentric orbit in the main asteroid belt where the mothership is ideally located in region of high density. Basic mission requirements associated with a Mother-Daughter architecture are established utilizing a relatively large number (10-20) daughter spacecraft distributed from a mothership within the asteroid belt for the purpose of executing sample and return missions.**

**A number of trade analyses are performed to establish system performance to changes in initial orbit, delta-V capability and maximum small spacecraft flight time. The balance between the initial delta-V burn and asteroid velocity matching are also examined, with a goal of minimizing the amount of fuel needed in the small spacecraft. Preliminary requirements for the system are established using these results, and a conceptual design is presented for comparison to other asteroid exploration techniques. Preliminary results indicate that the aforementioned concept of a mothership with small spacecraft is viable and should be considered as an alternative approach to first order surveying of the asteroid belt.**

## I. Introduction

They are over a half-million asteroids catalogued in the beltway region between Mars and Jupiter. These asteroids are diverse in their composition and vary in size from several thousand kilometers to few meters in size. Many are remains of proto-planets that encountered accretions or collisions. These bodies are like a giant jigsaw puzzle that holds insight into the formation of the early solar system and could hold vital clues to the locations of high value resources. New approaches are needed to establish a statistically significant sample space of chemical composition for the purpose of determining their overall makeup.

Experience from Hayabusa I and II [4], and Osiris-Rex show that sending a big spacecraft to rendezvous and get into orbit around these small asteroids is major challenge due their low gravity. These spacecraft need to constantly expend fuel. It has also been demonstrated that it is relatively difficult to use a single spacecraft to visit a large number of asteroids because of the time and delta-V required to transit between them. Better approaches are needed

Use of small-spacecraft based upon CubeSat technology to rendezvous, orbit and land on asteroids is arguably more efficient than using a large spacecraft. However, CubeSats propulsion systems are limited in terms of thrust, delta-V and lifespan. The mother-daughter architecture offers a possible solution where a large mother ship carries multiple nano-spacecraft to a central orbit within the asteroid belt, and then deploying them upon detection of a suitable asteroid. The nano-spacecraft then disperse to rendezvous with individual asteroids and then return to the mothership with samples. The mothership then analyzes the return samples for chemical composition and returns the results to earth. The concept is that the individual nano-spacecraft engage only with the mothership, reducing requirements on communication, delta-V and lifespan. This paper examines the effectiveness of such an architecture. The mothership is ideally located in region of high asteroid density. Delta-v and flight time requirements for nanosat sized spacecraft are then derived and compared to existing Nano-Satellite capabilities to determine feasibility, and a notional system is described.

---

[1] Ph.D. Student, Aerospace and Mechanical Engineering.
[2] Assistant Professor, Aerospace and Mechanical Engineering.
[3] Professor, Lunar and Planetary Laboratory

## II. Related Work

Typical mother-daughter architectures envision a fully capable mother-craft with a limited daughter-craft such as a CubeSat or small-sat being sent on a high-risk, high-reward short duration, mini-mission. Such concepts are not new and can be traced back to the Apollo 15 and 16, when a "Subsatellite" were launched on independent orbital missions around the moon [8]. The Apollo 15 "Subsatellite" lasted less than 2-weeks having entered an unstable lunar orbit, while Apollo 16's lasted more than a year having entered a "frozen orbit" and transmitted valuable data from lunar orbit and better insight into stable lunar orbits [9]. Another example is the Galileo probe, a 340 kg probe that achieved atmospheric entry into Jupiter and lasted nearly 1 hour, having survived some time under a 21-bar ambient pressure [10-11]. The probe likely melted in the super-critical hydrogen atmosphere. The probe determined Jupiter was denser and hotter than expected [10]. In other scenarios, in a mission like AAReST multiple daughter-crafts work together to form a telescope and can be readily be rearranged to provide multiple configurations [7]

A Keck Institute for Space Studies Report introduced several mother-daughter architectures and mission concepts [12]. This includes RELIC, which consists of 30+ 3U CubeSats to perform imaging and monitor energy transport from black holes. Others include measurement of reionization in the universe using a constellation of ESPA class small satellites. Another includes the Ionosphere Magnetosphere Coupling Constellation (IMCC) with plans to use 60 nanosatellites. The Solar Polar Constellation consists of a 6-12 CubeSat constellation to monitor variability of solar-effects. ExCSITE is another concept that utilizes multiple CubeSats and small-satellites including impactors for characterizing Europa's surface via imaging, gravity field mapping and chemical characterization. C/entinel, a concept to use multiple flyby and in-situ landers to perform proximity operations around small bodies and Lunar Cube Vibrations, that includes multiple fly-by and in-situ landers deployed as CubeSats with seismometers, thermal sensors and magnetometers to characterize the lunar surface.

NASA Goddard has pursued a concept for swarm exploration of the asteroid belt for several years. The concept called ANTS (for Autonomous Nano-Technology Swarm) envisions a large number (100 to 1000) of specialized nano-spacecraft dispersing through the asteroid belt to enable mass exploration [3]. Another area of interest involves proximity operations around small asteroids similar to the Hayabusa mission, providing an existing example regarding maneuvering and contact with small asteroids. Trajectory work is also done within the context of the Global Trajectory Optimization Competition [5], a yearly worldwide competition in orbital dynamics which has included asteroid visitation challenges in the past. Another proposed interplanetary spacecraft is the Hummingbird jointly proposed by NASA Ames and Microcosm [6]. Hummingbird is a spacecraft architecture intended to tour asteroids. It includes slots to carry CubeSats that would be deployed upon rendezvous with a target of interest, in-addition, it includes a telescope to observe an asteroid target at a distance. Other uses for swarms of small-spacecraft includes cooperative imaging and in [13-14], a fleet of small-satellites or CubeSats would be timed to flyby a small body to perform mapping. This saves fuel and time from having to get into orbit around a low-gravity small-body.

## III. Method

Performance for four stages of flight are established. Fuel usage statistics are gathered for (1) initial flyout, (2) braking to stop at the asteroid, (3) returning to the mothership, and (4) braking to rendezvous with the mothership. An overview of this concept is shown in Figure 1. The result is a statistical study of plausible asteroid sample and return missions where nano-spacecraft fly out to an asteroid, gather a small sample and return to the mothership for compositional analysis. Statistics on detection ranges are also gathered to establish requirements for an onboard optical search and detect sensor. Results are compared at various points around the belt to establish some understanding of opportunity variations as the mothership is placed in different orbits.

The distribution of asteroids within the main belt shows regions of relative high density which can be used as a starting point for low delta-V exploration (Nesvorny et al, 2015). The concept is to place a relatively large spacecraft mothership into such an orbit, where it will scan local space to establish the orbits of nearby asteroids, and then to deploy several nano-spacecraft to visit them. Such a mission could be tailored to explore specific asteroid types or families given their predominance in certain orbits. Each small spacecraft has the goal of visiting a single passing asteroid, matching velocities, retrieving a sample and returning to the mothership where the sample can be processed for chemical analysis.

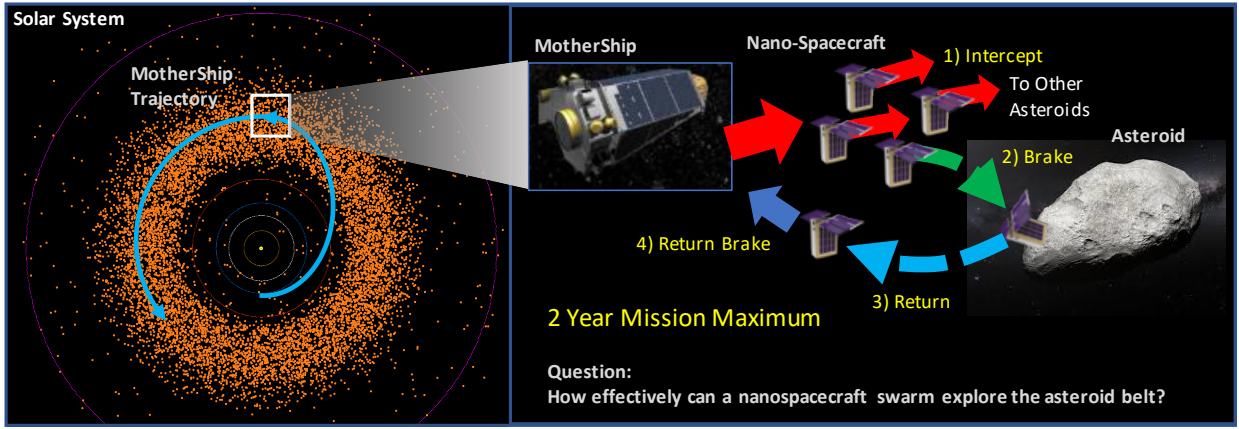

**Figure 1.** Mission Overview: Multiple Nano-spacecraft are deployed from a mothership within the asteroid belt, each executing a sample and return from a known asteroid.

Before a more in-depth feasibility analysis was conducted, work was done to establish the plausibility of the basic concept. Because the lifetime and delta-v capabilities of individual nano-spacecraft are limited, it is reasonable to question whether such a spacecraft could have the capability to even reach other asteroids from the mothership orbit. Figure 2 gives an example of a maneuver boundary defined from a modest, isotropic 300 m/s initial burn. The resulting kinematic volume starts off as a growing sphere, but over time stretches into an ellipsoid in line with longer term orbital dynamics. the resulting kinematic volume is large enough to contain 14 asteroids after a 1-year propagation. Careful selection of the initial mothership orbit may provide even higher densities.

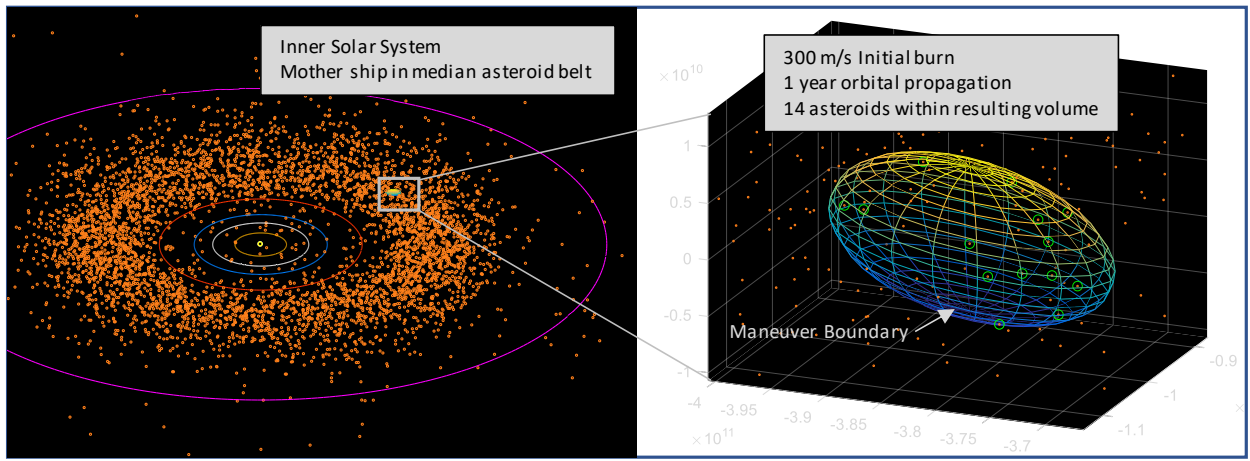

**Figure 2.** Intercept is established by determining inclusion within a nano-spacecraft maneuver boundary

Analysis of this architecture is accomplished via statistical analysis of simulation output. The orbital ephemeris of all cataloged asteroids is loaded, and the mothership is notionally placed in a high density orbit near the center of the belt as shown in Figure 3. The simulation then establishes the kinematic volume which the nano-spacecraft can reach by examining the effect of an isotropic velocity step applied at the moment of deployment, and then propagating the resulting kinematics over time. This volume, initially sphere shaped, grows and stretches over time consistent with orbital dynamics, and statistics are gathered on the number of cataloged asteroids which pass within.

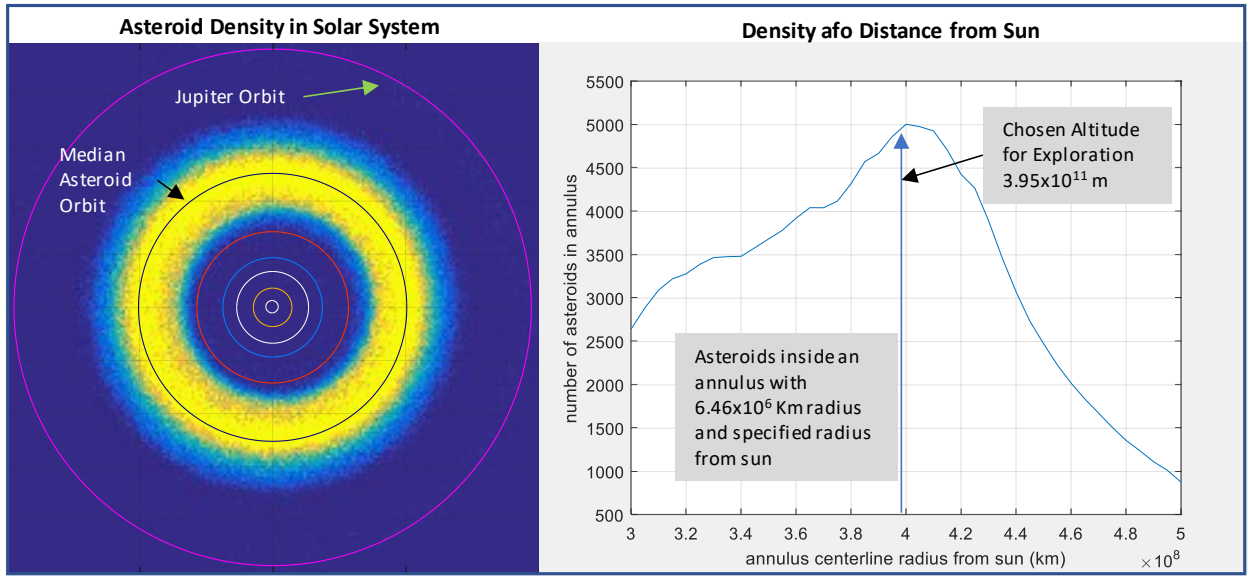

**Figure 3.** Nominal Mothership orbit is placed at the location of highest asteroid density

A MatLab simulation is constructed which loads the NASA/JPL database of approximately 518,000 cataloged asteroids, establishes initial positions and velocities for each, and then propagates their orbits using standard time integration. Amongst them, the mothership is then placed in a high density candidate orbit (as shown in Figure 3) and also propagated. At a defined time zero, the simulation then creates a notional group of 400 nano-spacecraft and gives each of them a relative outgoing velocity vector of 1500m/s relative to the mothership using a 20x20 latitude/longitude grid. These trajectories, which describe a sphere in the short term, are then propagated along with mothership and asteroids for the period of 1.5 years, with the presumption that a two year total lifespan for the nanosatellite as a reasonable limit on their capability.

As the nano-spacecraft fly out, the spherical boundary described by their position grows and elongates as per orbital dynamics, and after 1.5 years (approximately one third of an orbit), the resulting ellipsoid representing the outer boundary of possible travel is approximately $4 \times 10^{11}$m long, enclosing over 8100 asteroids as shown in Figure 4.

At each stage in the propagation, if an asteroid is within the ellipsoid of the kinematic capability, interpolation is done with respect to the outer boundary to establish the delta-V necessary to reach the asteroid at that particular time. This is stored as a candidate sample and return trajectory. Typically as the asteroid enters the maneuver boundary, the required delta-V for intercept is higher, decreasing to a minimum as more time is allowed for the intercept. The time dependency of the required delta-V for intercept is shown in the left side of Figure 5. Braking delta-V is established by simply establishing the relative velocity of the nano-spacecraft w/r to the asteroid. The resulting total delta-V for the outbound journey is limited to 3.0 km/s to prevent the number of candidate missions from becoming too large.

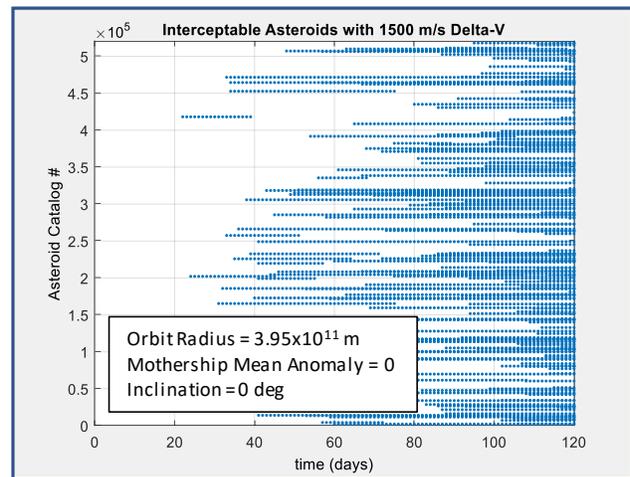

**Figure 4.** More asteroids can be reached as nano-spacecraft flyout time increases

Returning to the mother ship then requires the execution of a 3-dimensional gradient (delta-v in x, y and z directions) combined with a golden section search algorithm [2] which determines the fuel necessary to return to the mothership exactly two years after launch, thus minimizing the total delta-V usage at the expense of using the entire expected lifetime of nano-spacecraft subsystems. Finally, the relative velocity of the returning spacecraft with respect to the mothership is calculated to establish the braking necessary to match velocities.

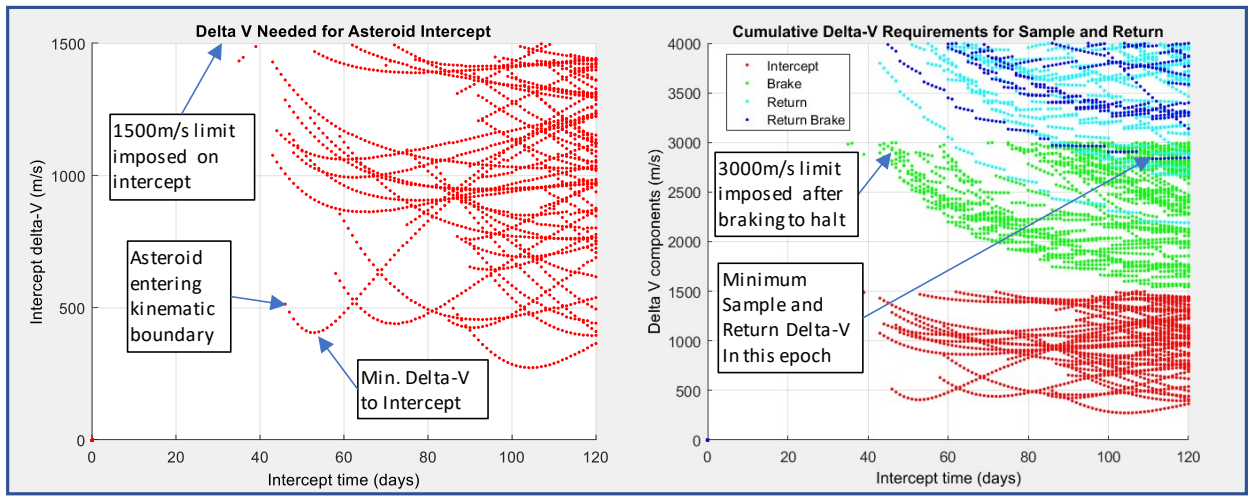

**Figure 5.** Overview of Delta-V components for Sample and Return from center of asteroid belt

These four components of delta-V are summed to establish the overall propulsive requirement for the nano-spacecraft as a function of the intercept time, shown cumulatively in the right frame of Figure 5. There are a very large number of candidate missions depending upon the amount of fuel that the nano-spacecraft can carry, and there are clearly viable missions for delta-Vs of 3000m/s and higher. In fact, if intercept possibilities are allowed all the way out to 1.5 years (out of a two year total nano-spacecraft lifetime), there are typically several opportunities available at even lower delta-V capabilities.

By keeping track of the delta-V requirements for each candidate asteroid, and then for each time in which they can be intercepted, overall statistics are established producing a curve establishing the number of asteroids which can be visited vs the amount of delta-V on board each nano-spacecraft. This process was repeated for 8 initial mean anomalies for the mothership to provide an understanding of how performance is affected by different initial mothership placement. Actual missions would require some unknown additional fuel budget for docking maneuver and navigation uncertainties, but no additional delta-V budget was established for these at this time.

## IV. Analysis/Results

The first step of the analysis is to establish which asteroids can be reached from the mothership, and how much delta-V is utilized to intercept them. The simulation propagates positions from initial delta-V's distributed spherically around the mothership at time zero. The resulting spherical kinematic boundary grows and slowly becomes ellipsoidal as it interacts with the sun's gravity. When an asteroid passes inside this defined kinematic boundary, it is flagged as an intercept opportunity, and then the relative distance of the asteroid with respect to the kinematic boundary is used to prorate down to the required delta-V for intercept at that particular time. The right side of Figure 5 shows the result of the process for the first 120 days of possible intercepts, showing how delta-V to intercept changes as a function of when the intercept occurs.

As per the methodology described above, the landing, return and docking delta-V's are then calculated for each intercept possibility at each simulation time step. Figure 6 shows a plot of the cumulative additional delta-V required for each of these steps, but now with intercepts occurring as late as 1.5 years into the nano-spacecraft lifetime. It is clear in looking at these results in Figures 5 and 6, that the first intercept opportunity is not necessarily the best from a fuel usage standpoint, and there exist a large number of intercept opportunities which are not viable for a sample and return mission due to excessive fuel requirements. It is interesting to note that the delta-V requirements for a simple intercept mission are so much smaller than the total for sample and return, that it might be worthwhile to discuss the possibility of an impactor/flyby spectrometer spacecraft pair as an alternate nano-spacecraft configuration.

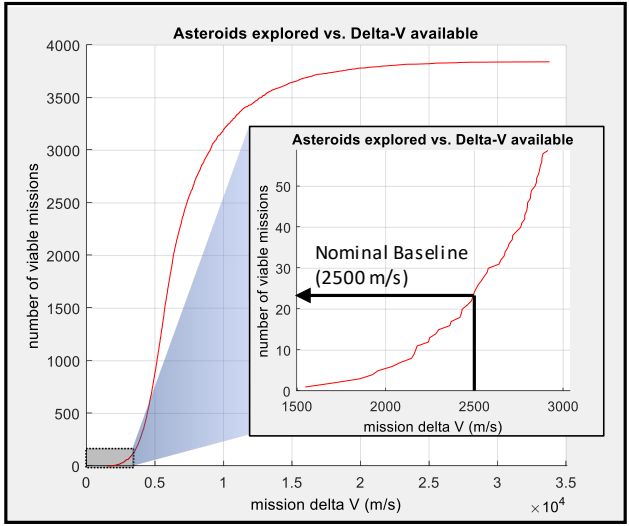

**Figure 6.** Number of viable missions rise quickly as nano-spacecraft delta-V increases above 2000m/s

The availability of missions as a function of delta-V could clearly be driven by the choice of initial position of the mothership within the asteroid belt, and as such, the above analysis was repeated for anomalies at 45 degree intervals around the chosen orbit. The minimum delta-V utilized for each candidate asteroid sample/return is sorted and then plotted as a distribution as shown in the larger frame of Figure 6. The advantage of this presentation is we get a reading of how many asteroids can be visited as a direct function of available nano-spacecraft delta-V.

To establish some understanding of the variation in results for different mothership orbits, the initial location of the mothership is placed at different initial orbital positions every 45 degrees around the asteroid belt within the initial circular orbit chosen. The simulation is then executed eight times and the cumulative statistics of delta-V and detection range are gathered and presented in Figure 7. Here it can be seen that there is significant variation of results with even this simple change in orbital position, and the viability of the mission can be significantly improved by a more detailed survey of intercept possibilities before choosing the final mothership orbit.

Finally, assuming that some amount of local steering control might be required of the mothership in order to effectively guide the nano-spacecraft to their target asteroids, distance statistics between the mothership and the target asteroids at time zero were established to determine the range requirement of a possible passive visible sensor. These initial ranges are shown as a function of the total delta-V required for each mission with the presumption that the chosen missions will generally have lower initial ranges. Figure 8 shows that initial ranges do in fact increase statistically as higher delta-V missions are included.

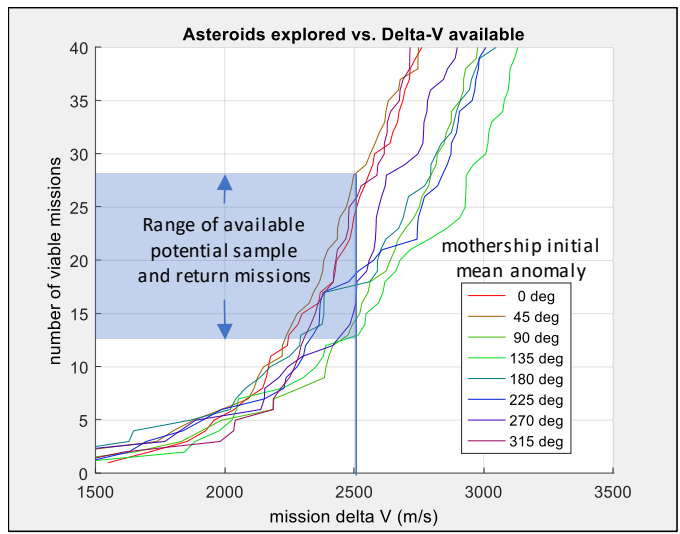

**Figure 7.** Although choice of initial orbit anomaly is important, 2.5 km/s enables a significant number of missions independent of initial position

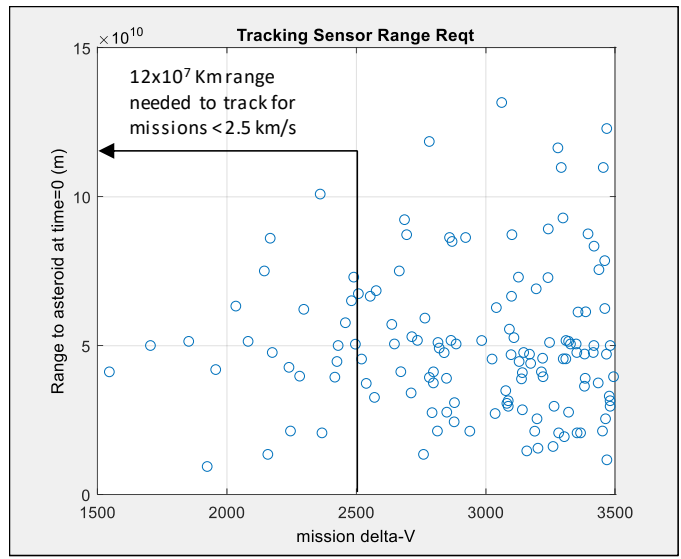

**Figure 8.** If Mothership must track asteroid Nano-spacecraft launch, about 120 million kilometers of range is required.

## V. Conclusions

The concept of placing a mothership in the middle of the asteroid belt and then deploying swarms of nano-spacecraft on individual sample and return missions to chosen asteroids is realistic and within the boundaries of plausible delta-V usage. Statistics show that careful choice of missions can result in a robust number of visited asteroids. Although the amount of delta-V required for the aforementioned mothership/nano-spacecraft system architecture is not insignificant, the cost of providing small spacecraft with on the order of 2.5 km/s delta-V is small compared to the cost of sending single larger spacecraft to do the same job sequentially. Each visited asteroid adds multiple kilometers/second of delta-V to the mission and the resulting spacecraft mass increases exponentially as the number of asteroids visited increases. The use of a nano-spacecraft swarm to execute these missions in parallel is a significantly more mass-effective technique.